\documentstyle[12pt,epsf]{article}
\def\lsim{\mathrel{\vcenter{\hbox{$<$}\nointerlineskip\hbox{$\sim$}}}}
\newcommand{\be}{\begin{equation}}
\newcommand{\ee}{\end{equation}}
\newcommand{\ba}{\begin{eqnarray}}
\newcommand{\ea}{\end{eqnarray}}

\def\21{$SU(2) \otimes U(1) $}

\def\lsim{\raise0.3ex\hbox{$\;<$\kern-0.75em\raise-1.1ex\hbox{$\sim\;$}}}
\def\gsim{\raise0.3ex\hbox{$\;>$\kern-0.75em\raise-1.1ex\hbox{$\sim\;$}}} 

\newcommand{\mx}{\left[\begin{array}}
\newcommand{\finmx}{\end{array}\right]} 
\newcommand{\mxp}{\left(\begin{array}} 
\newcommand{\finmxp}{\end{array}\right)} 
\def\beq{\begin{equation}}
\def\eeq{\end{equation}}
\def\bea{\begin{eqnarray}}
\def\eea{\end{eqnarray}}

\def\mathbf#1{\hbox{\bf #1}}
\def\textrm#1{\hbox{#1}}

\def\lsim{\raise0.3ex\hbox{$\;<$\kern-0.75em\raise-1.1ex\hbox{$\sim\;$}}}
\def\gsim{\raise0.3ex\hbox{$\;>$\kern-0.75em\raise-1.1ex\hbox{$\sim\;$}}}
\newcommand {\ignore}[1]{}

\begin{document}
\vspace*{-1in}
\renewcommand{\thefootnote}{\fnsymbol{footnote}}
\begin{flushright}
\texttt{
} 
\end{flushright}
\vskip 5pt
\begin{center}
{\Large{\bf Generalized Bounds on Majoron-neutrino couplings}}
\vskip 25pt 

{\sf R. Tom{\`a}s},
{\sf H.  P\"{a}s \footnote{New address:
Department of Physics and Astronomy, Vanderbilt University,
Nashville, TN 37235, USA}},
{\sf J. W. F. Valle}
\vskip 10pt
{\small Institut de F\'{\i}sica Corpuscular - C.S.I.C., 
Departament de F\'{\i}sica Te{\`o}rica - Univ. de Val{\`e}ncia \\
Edifici Instituts d'Investigaci{\'o} - Apartat de Correus 2085 - 46071 Val{\`e}ncia, 
Spain}\\

\vskip 20pt

{\bf Abstract}
\end{center}

\begin{quotation}
{\small 
  
  We discuss limits on neutrino-Majoron couplings both from laboratory
  experiments as well as from astrophysics. They apply to the simplest
  class of Majoron models which covers a variety of possibilities
  where neutrinos acquire mass either via a seesaw-type scheme or via
  radiative corrections.  By adopting a general framework including CP
  phases we generalize bounds obtained previously. The combination of
  complementary bounds enables us to obtain a highly non-trivial
  exclusion region in the parameter space. We find that the future
  double beta project GENIUS, together with constraints based on
  supernova energy release arguments, could restrict neutrino-Majoron
  couplings down to the $10^{-7}$ level.  }

\end{quotation}

\vskip 20pt  

\setcounter{footnote}{0}
\renewcommand{\thefootnote}{\arabic{footnote}}

\section{Introduction}

The confirmation of the zenith--angle--dependent atmospheric neutrino
deficit by the Superkamiokande experiment generally has been
understood as first significant hint for neutrino masses and thus
particle physics beyond the standard model \cite{skatm00}.
The other long-standing puzzle of particle physics is the deficit of
solar neutrinos \cite{solar}.
Altogether they constitute the most important milestone in the search
for phenomena beyond the Standard Model (SM), indicating the need for
oscillations involving all three active neutrino
species~\cite{Gonzalez-Garcia:2001sq}.
The mounting experimental activity in this field promises a bright
future for neutrino physics which may prove to be a most valuable
source of information on the structure of a more complete theory
underlying the standard model of particle physics.

An elegant way to introduce neutrino masses is via the spontaneous
breaking of an ungauged lepton number symmetry through a non-zero \21
singlet vacuum expectation value (VEV) of a scalar field.
This may be implemented in conventional~\cite{models1,models1rad} as
well as supersymmetric models ~\cite{models2}.
The couplings of the corresponding Goldstone boson, generically called
Majoron and denoted by $J$, are rather model-dependent~\cite{val91}.
Here we consider the simplest class of Majoron models, where the
Majoron-neutrino coupling matrix $g_{ij}^M \propto m_{ij}$ is
proportional to the neutrino mass matrix~\cite{Schechter:1982cv}, so
that in the mass eigenstate basis diag~$(m_1,m_2,m_3)$ the Majoron
neutrino couplings $g_i$ are diagonal, to lowest order approximation
~\footnote{This proportionality may be avoided in more complex models,
  such as those in ref.~\cite{mod3,mod4}},
\be 
g_{ij}^M \simeq \delta_{ij} g_i.
\label{gdiagonal}
\ee

This covers a variety of possibilities including both
seesaw-type~\cite{models1} as well as radiative
models~\cite{models1rad}.

Limits on this quantity obtained from laboratory experiments searching
for Majoron-emitting pion or kaon decays are rather weak, with the
exception of double beta decay \cite{hpkk}. On the other hand Majoron
emitting neutrino decays affect the expected neutrino luminosity and
spectra which are constrained by the observed signal from SN1987A,
providing stringent restrictions~\cite{kach}.  While the limits of
laboratory experiments on rare decays are given in the weak basis,
bounds from processes in supernovae occur in a dense medium and are
expressed in the medium eigenstates (see below). In the present work
we discuss the correlations of the different limits and their
translation into the mass basis, extending the earlier paper ref. \cite{kach}.
In the next section we derive the
expressions for medium and weak eigenstates, following
\cite{kach,giunti}.  In section 3 we review the bounds obtained from
the supernova SN1987A using various considerations. In contrast to
ref.~\cite{kach} here we include the study of the effects associated
to the Majorana CP violating phases present in theories of massive
neutrinos~\cite{Schechter:1980gr,Schechter:1981gk}. Moreover we
investigate (Section 4) the recent bounds from neutrinoless double
beta decay as well as those that could be attained in future
experiments such as GENIUS~\cite{genius}.  The resulting exclusion
plots are discussed in section 5 in the mass basis.

\section{Neutrino mixing in three bases}

For neutrinos propagating through a medium one has to deal with three
kinds of eigenstates: Flavor eigenstates $\nu_{\alpha}$, mass
eigenstates $\nu_i^{(h_i)}$ with masses $m_i$, and, depending on the
environment, medium eigenstates $\tilde{\nu}_i^{(h_i)}$.  The flavor
eigenstates are defined as
\be
\nu_{\alpha}= \sum_i U_{\alpha i} \nu_i^{(h_i)} 
\ee
and the medium 
eigenstates are
\be
\tilde{\nu}^{(h_i)}_i= \sum_j \tilde{U}_{ij} \nu_j^{(h_j)}. 
\ee
Here the superscript $(h_i)=\pm 1$ refers to the helicity of the
state.  In the general case it is impossible to diagonalize
simultaneously the mass and potential terms. Thus one has to solve the
field equations in detail.  In a two-component field formalism, where
a left-handed four-component field $\nu$ expressed in the chiral
representation of the $\gamma$ matrices \cite{Schechter:1980gr} is
related to the corresponding two-component field $\phi$ by
$\nu_L^T=(\phi^T,0)$ \cite{giunti} \footnote{The notation here
  coincides to the one of \cite{val91,Schechter:1980gr} up to a factor
  of $i$.}  , the Lagrangian can be written in the mass basis as
\bea
 \mathcal{L}_{\rm tot} &=& \mathcal{L}_{\rm 0} + \mathcal{L}_{\rm med} + 
\mathcal{L}_{\rm int} \nonumber
\\ &=&
   \sum_i \phi_i^\dagger(i\partial_0 -i {\vec{\sigma}} \times 
{\vec{\nabla}})\phi_i -\frac{m_i}{2}(\phi_i^T i\sigma_2 \phi_i -
\phi_i^\dagger i\sigma_2 \phi^*)
   + \sum_{i,j}   \phi_i^\dagger  V_{ij}  \phi_j -\nonumber \\
& &   - J \sum_{ij} g^M_{ij}(\phi_i^T \sigma_2 \phi_j + \phi_i^\dagger \sigma_2
\phi_j^*)   
\eea
where $\sigma_2$ and $ \vec{\sigma}$ denote Pauli matrices. Here the
free Langrangian ${\mathcal L}_0$ and $\mathcal{L}_{\rm med}$ describe
the propagation in vacuo and the effects of matter described by the
potential matrix $V_{ij}$, respectively, whereas $\mathcal{L}_{\rm
  int}$ takes into account the presence of neutrino-Majoron
interactions which may lead to decays.  One has now to consider the
decays $\tilde{\nu}_i^{h_i}\!(p_i)\to
\tilde{\nu}_j^{h_j}\!(p_j)+J(q)$, where $\tilde{\nu}_i^{h_i}\!(p_i)$
and $\tilde{\nu}_j^{h_j}\!(p_j)$ are energy--eigenstate Majorana
neutrinos that propagate in matter with four-momenta
$p_i=(E_i^{h_i},{\vec{p}_i})$ and $p_j=(E_j^{h_j},{\vec{p}_j})$, and
helicity $h_i$ and $h_f$ respectively.  In order to obtain these
energy--eigenstates one has to take $\cal{L}_{\rm{0}}+\cal{L}_{\rm
  med}$ and calculate the resulting field equations.
\beq
(i\partial_0-i \vec{\sigma}  \times  \vec{\nabla})\phi_i(x)+m_ii\sigma_2\phi_i^*
-\sum_{j=1}^N V_{ij}\phi_j(x)=0.
\label{eqmotion}
\eeq
One solves these field equations by expanding the fields $\phi_i(x)$ as
superpositions of plane-wave spinors with definite helicity,
\cite{giunti,ma88},

\bea \phi_i(x)=\int\frac{d\vec{p}}{\sqrt{(2\pi)^3}}e^{i\vec{p} \times 
  \vec{x}} \left \{ [
  P_\alpha^i(\vec{p},t)+N_\alpha^i(\vec{p},t)]\alpha(\vec{p})
  +[P_\beta^i(\vec{p},t)+N_\beta^i(\vec{p},t)]\beta(\vec{p}) \right \}
\eea where $\alpha(\vec{p})$ and $\beta(\vec{k})$ are helicity
eigenstates and $P^i$ and $N^i$ denote positive and negative frequency
components of the field under consideration. One should now substitute
this expression in the equations (\ref{eqmotion}), whose
diagonalization would give rise to the desired eigenstates. It can be
shown, though, that for relativistic neutrinos the positive-frequency
components decouple from the negative-frequency ones and the energy
eigenstates obtained in this way result to be the same as those
obtained from the diagonalization of the usual MSW equation
\cite{giunti}, which can be stated as
\beq
 i\partial_t \nu_i^{(h)} = (H^{\rm rel}_{ij}+U_{i\alpha} V_{\alpha \beta} 
U^\dagger
_{\beta j})
\nu_j^{(h)}.
\eeq
Here
$ H^{\rm rel}_{ij} \approx (p+m_i^2/(2p))\delta_{ij} $ and
$V_{\alpha\beta}$ is the potential matrix in the weak basis,
\beq
V_{\alpha\beta} = \left(\begin{array}{ccc} V_C+V_N & 0 & 0\\ 0 & V_N &
  0 \\ 0 & 0 & V_N\end{array}
\right) \,.
\eeq
The potentials induced by the charged and neutral currents are $V_C =
\sqrt{2} h G_F n_B ( Y_e + Y_{\nu_e})$ and $V_N = \sqrt{2} h G_F n_B
\left( -\frac{1}{2}Y_N + Y_{\nu_e} \right)$, where $Y_i=(n_i-n_{\bar
  i})/n_B$ and $n_B$ is the baryon density. Diagonalizing
$H^{\rm rel}+U V U^\dagger $ yields the medium eigenstates
$\tilde{\nu}_i^{(h)}=\tilde{U}_{ij}^{(h)} \nu_{j}^{(h)}$.

\begin{table}[!t]
\begin{center}
 \begin{tabular}{c|c|c}
 medium state & weak state & potential \\ \hline
$\tilde{\nu}_1^{(+)}$ & $\bar{\nu}_e$ &$ -(V_C+V_N)$\\
$\tilde{\nu}_2^{(+)}$ & $\bar{\nu}_{\mu'}=
c_{23}\bar{\nu}_\mu-s_{23}\bar{\nu}_\tau$ &$ -V_N$\\
$\tilde{\nu}_3^{(+)}$ & $\bar{\nu}_{\tau'}=
s_{23}\bar{\nu}_\mu+c_{23}\bar{\nu}_\tau$ 
&$ -V_N$\\
$\tilde{\nu}_1^{(-)}$ & $\nu_{\mu'}=c_{23}\nu_\mu-s_{23}\nu_\tau$ &$ V_N$\\
$\tilde{\nu}_2^{(-)}$ & $\nu_{\tau'}=s_{23}\nu_\mu+c_{23}\nu_\tau $ &$V_N$\\
$\tilde{\nu}_3^{(-)}$ & $\nu_e$ & $V_C+V_N$\\
 \end{tabular}
\end{center}
\caption{Medium eigenstates $\tilde{\nu}_i^{\pm}$, equivalent weak
eigenstates in the limit $|V_{\alpha\alpha}|\gg m_i^2/(2p)$ 
and their potential energy. The rotation in the $\nu_\mu-\nu_\tau$ subspace
is parametrized by $c_{23}=\cos \theta_{23}$ and $s_{23}=\sin \theta_{23}$,
where the arbitrary argument has been chosen to coincide with the
mixing angle $\theta_{23}$ in vacuo
(from \protect{\cite{kach}}).}
\label{states}
\end{table}

In the three-flavor neutrino case the mixing matrix $U$ can be
parametrized as $U=U_{23}U_{13}U_{12} U_0$, where the matrices
$U_{ij}= U_{ij}(\theta_{ij})$ perform the rotation in the $ij$-plane
by the angle $\theta_{ij}$ and $U_0$ includes possible CP-violation
effects~\cite{Schechter:1980gr,Schechter:1981gk}. In the following we
will assumme $\theta_{13}=0$, motivated both by detailed fits of the
present solar and atmospheric neutrino
anomalies~\cite{Gonzalez-Garcia:2001sq} as well as by the reactor
results of the Chooz experiment~\cite{chooz}.  This simplifies the
mixing matrix to $\nu_\alpha=U_{\alpha i}\nu_i= U_{23}U_{12}U_0\nu_i$
\cite{Schechter:1980} and allows us to set $\theta_{12}=\theta_\odot$
and $\theta_{23}=\theta_{\rm atm}$.  Now for light neutrinos near the
neutrinospheres in supernovae the condition $|V_{\alpha\alpha}|\gg
m_i^2/(2p)$ holds and, since in the weak basis the potential is
diagonal, the medium states can be identified with the weak ones up to
an arbitrary rotation in the $\nu_\mu-\nu_\tau$ subspace.  In order to
simplify the expressions we exploit this freedom by choosing this
arbitrary rotation angle to coincide with $-\theta_{23}$, see Table
\ref{states}. This allows us to identify the coupling matrix in the
medium basis with the one in the weak basis up to the rotation
\bea
\tilde{g}_{ij} = U(-\theta_{23})g^W_{\alpha \beta} U^T(-\theta_{23})\equiv 
g_{\alpha'\beta'}.
\eea
Taking now into account that $g^M = U^T~g^W~U$ and substituting the
explicit expressions for the $U_{\alpha i}$ matrices relating mass and
weak eigenstates, one gets the following expression, $ \tilde{g}_{ij}
= U_{12}U_0^*~ g_{ij}^M~ U_0^\dagger U_{12}^T $, or explicitly
\bea
\tilde{g}= \left( \begin{array}{ccc}
g_{ee}  &       g_{e\mu'}       &       g_{e\tau'}\\
g_{\mu'e}&      g_{\mu'\mu'}    &       g_{\mu'\tau'}\\
g_{\tau'e}&     g_{\tau'\mu'}   &       g_{\tau'\tau'}
\end{array} \right)
=
\left( \begin{array}{ccc}
g_1\cos^2\theta_\odot+g_2\sin^2\theta_\odot~e^{-2i\delta} &
\frac{1}{2}(-g_1+g_2~e^{-2i\delta})\sin 2\theta_\odot   &       0\\
\frac{1}{2}(-g_1+g_2~e^{-2i\delta})\sin 2\theta_\odot  & 
g_1\sin^2\theta_\odot+g_2\cos^2\theta_\odot~e^{-2i\delta} & 0\\
0 & 0 & g_3
\end{array} \right).
\label{gtildematrix}
\eea
This choice of the rotation in the $\nu_\mu-\nu_\tau$ subspace leads
to a relation between medium and mass eigenstates characterized only
by the solar angle $\theta_\odot$ and by the Majorana CP violating
phase $\delta$~\cite{Schechter:1980gr,Schechter:1981gk}. Using the
definitions $\Delta m_{12}^2 =\Delta m_\odot^2$ and $\Delta
m_{23}^2=\Delta m_{atm}^2$ together with the assumptions in
\ref{gdiagonal} one can easily translate the bounds obtained in the
weak or medium basis into the mass basis and, in addition, express
them in terms of only two independent parameters, for instance
$(m_1,g_1)$ via
\bea
\label{parametrization}
g_2=g_1 \sqrt{1+\frac{\Delta m_{\odot}^2}{m_1^2}}~~,~~~~~
g_3=g_1 \sqrt{1+\frac{\Delta m_{\odot}^2+\Delta m_{atm}^2}{m_1^2}}~.
\eea

\section{Supernova bounds}

There is a variety of different arguments based on supernova physics
which lead to restrictions on neutrino properties.  Processes
involving Majoron-neutrino couplings may prevent a successful
explosion as well as substantially affect the observed neutrino
spectra.  A crucial feature to notice is that the effective mass
induced by the interactions of neutrinos with background matter breaks
the proportionality between the neutrino mass matrix and the
neutrino-Majoron coupling matrix $g_{ij}^M$.  This follows from the fact
that the thermal background in the supernova environment consists only
of particles of the 1st generation, thus distinguishing the electron
flavour from the others.  We now describe three different arguments
used \cite{kach} in order to restrict the relevant parameters.

\subsection{Constraints from Neutrino spectra}

The idea behind this bound is that Majoron--induced transitions
between the neutrino flavors could change the energy spectra of the
single flavors.
At the typical temperatures of the SN core $\nu_{\mu,\tau}$ only
interact with the medium via neutral currents giving rise to a smaller
cross section than that corresponding to the electron (anti)neutrinos,
which feel both neutral and charged currents.
Since the opacity of the heavier $\nu_{\mu,\tau}$ flavors is smaller
than for the $\nu_e$, their energy-exchanging reactions freeze out in
the denser region of the protoneutron star, leading to lower spectral
temperatures of $\nu_e$ compared to $\nu_{\mu,\tau}$. This expected
spectrum can be distorted due to the decays $\tilde{\nu}_i^\pm \to
\tilde{\nu}_j^\mp + J$.  Besides the effects of such decays one has to
keep in mind the possible oscillation which neutrinos could undergo
along their journey to the Earth. In order to consider both aspects we
have defined the effective survival probability as 
\bea
N=N_{decay} \times  N_{osc}, 
\eea
where $N_{decay}$ stands for the survival probability of a
$\tilde{\nu}_i^\pm$ emitted from its energy sphere and can be computed
as
\bea
N_{decay}(\tilde{\nu}_i^\pm)\simeq 
\exp \left\{ -\int_{R_{E,\tilde{\nu}_i^\pm}}^
\infty 
dr\sum_{j}\Gamma(\tilde{\nu}_i^\pm
\to \tilde{\nu}_j^\mp+J) \right\}.
\eea
Within our relativistic approximation, the helicity-flipping neutrino
decays rate are given by
\bea
\Gamma(\tilde{\nu}_i^\pm \to \tilde{\nu}_j^\mp+J)=\frac{|\tilde{g}_{ij}|^2}
{16\pi}(V_i-V_f).
\eea
Coming to the oscillation term, the corresponding neutrino survival
probability $N_{osc}$, will depend on the neutrino mixing angle and
squared mass difference.  We will analyse separately the different
solutions of the solar neutrino problem namely small-angle MSW
(SMA-MSW), large-angle MSW (LMA-MSW), LOW-MSW and the just-so case.
Details about the present status and required parameters of the
various solutions can be found in the global analysis of neutrino data
presented in ref.~\cite{Gonzalez-Garcia:2001sq}. Such a study favours
a rather small value for the angle $\theta_{13}$, mainly because of
data from reactors~\cite{chooz}.
In the first three cases neutrinos will propagate through the
supernova environment adiabatically.  Therefore they will emerge as
energy eigenstates, which in vacuum coincide with the mass
eigenstates, without any oscillation occuring on the way from the SN
to Earth. If one takes into account that neutrinos have to traverse a
distance, $d$, of matter in the Earth to reach the detectors,
Kamiokande and IMB, one gets the following expression for their
survival probability \cite{BSS,JNR},
\bea
N_{osc}=1-\{\sin^2\theta_\odot -\sin 2\theta_m~\sin(2\theta_\odot-2\theta_m)
\sin^2 \left(\frac{\pi d}{l_m}\right)\},
\eea
where $l_m$ and $\theta_m$ denote the oscillation length and the
mixing angle in matter, respectively.
As has been previously noted for the simplest choice $\theta_{13}=0$
one has that, besides the fact that $\nu_\mu$ and $\nu_\tau$ behave
the same way in the supernova, the conversion $\nu_e \to \nu_{\mu'}$
will be the only oscillation involving electron (anti)neutrinos,
allowing us to set the angle which characterizes their mixing to
$\theta_\odot$.

In the vacuum solution case 
the neutrinos emerge essentially as flavor eigenstates which then
oscillate on their way to Earth. Therefore one has 
\bea
N_{osc}=1-\frac{1}{2}\sin^22\theta_\odot.  
\eea
In order to get information on the coupling constants we will
conservatively require that at least half of the initial electron
antineutrinos emerging from the SN1987A survive, $N>0.5$, accounting
for the rough agreement between the expected and the detected SN1987A
signals. In order to analyse the implications of this restriction one
must generalize the simplest argument used in \cite{BSS} since
neutrinos may loose energy as a result of majoron decays.

This allows us to get some limits on the coupling parameter of the
order of $g_1(g_2)\leq \mathrm{few} \times  10^{-4}$ from the first three
solutions to the solar neutrino problem.  For the case of vacuum
oscillations, though, the solution is already disfavored by the
SN1987A data even in the absence of neutrino decays \cite{BSS}.
Though they may narrow it down considerably, the above arguments do
not totally close the allowed window of neutrino-Majoron couplings,
neither for the SMA, LMA nor LOW solutions, even for a supernova in
our milky way.

\subsection{Constraints from Majoron luminosity}

This bound is based on the observation that neutrino decays into
Majorons could supress the energy release contained in the neutrino
signal.  Under the assumption of small $\nu_e-\nu_x$ mixing the
neutrino signal observed in SN 1987A is in good agreement with
numerical computations of the total binding energy released in a
supernova explosion.  An analysis of decay and scattering processes
involved yields the exclusion region  \cite{kach}
\be{}
3 \times 10^{-7} < |\tilde{g}_{ij}| < 2 \times 10^{-5}.
\label{lbound}
\ee
For $|\tilde{g}_{ij}|$ values smaller than $3 \times 10^{-7}$ the
Majoron neutrino coupling becomes too small to induce any effect.
On the other hand for $|\tilde{g}_{ij}| > 2 \times 10^{-5}$ Majorons get
trapped in the core and do not contribute to the energy release.

Another point to observe is that CP violating phases affect these
limits.  This follows from the appearance of the phase $\delta$ in
explicit form of the Majoron neutrino coupling constants given in
eq.~(\ref{gtildematrix}).
In order to eliminate such an explict CP phase dependence when
translating the limit on $|\tilde{g}_{ij}|$ into the mass basis we have
analyzed for each term of eq.  (\ref{gtildematrix}) the excluded
region for different values of $\delta$ and subsequently considered
the intersection of the resulting excluded regions.  This conservative
procedure allows us to rule out part of the parameter space
irrespective of the value of the CP phase.

As an example we illustrate in Figs. 1 and 2 the regions excluded for
the LMA MSW solution to the solar neutrino problem. The luminosity
bound can be described in two steps.  In the first one we take one
$\tilde{g}_{ij}$ from eq.  (\ref{gtildematrix}) and by means of eq.
(\ref{parametrization}) we write it in terms of $g_1-m_1$. This way we
obtain an expression for the energy loss which depends explicitly upon
the CP phase $\delta$.  Now, by varying that CP phase the bound given
in eq. (\ref{lbound}) is translated into different ruled out regions.
We show in Fig. 1 the resulting bound on $|g_{ee}|$ assuming two extreme
cases, $\delta = 0$ (solid lines), and $\delta = \pi/2$ (dotted lines).
Notice that for the latter case the bound disappears because of a
cancellation between the two terms in $|g_{ee}|$. 
In order to remove the phase we therefore consider the intersection as
the most conservative choice.

Now turn to the implications of the luminosity bound to the other
components of the Majoron neutrino coupling matrix elements.
Once we have obtained those intersecting regions for each
$\tilde{g}_{ij}$ we simply take the union of them, giving rise to a
final highly non-trivial exclusion region, as can be seen in Fig. 2.

It is important to notice that the shape of such regions is
characterized by the values of the square root of $\Delta m^2_\odot$
and $\Delta m^2_{atm}$. Let us first consider $\tilde{g}_{ij}$ with
$i,j \neq 3$. In this case only $\Delta m^2_\odot$ appears in eq.
(\ref{parametrization}), so that for $m_1 \gg \Delta m^2_\odot$ one has
$g_2 \sim g_1$ giving rise to a vertical line with no dependence on
$m_1$, as noted in the figures. In contrast, for $m_1 \ll \Delta
m^2_\odot$ one has $g_2 \sim g_1 \frac{\sqrt{\Delta m^2_\odot}}{m_1}$
with a explicit dependence on $m_1$ which strengthens the bound for
lower $m_1$ values.
Let us now consider the limit coming from $g_3$.  In this case the
characteristic mass scale is always given by $\Delta m^2_{atm}$, eq.
(\ref{parametrization}), irrespective of the particular solutions to
solar neutrino problem that one may wish to consider. As a result, for
the LOW and (quasi)--vacuum cases the difference between $\Delta
m^2_{atm}$ and the solar mass scale $\Delta m^2_\odot$ is so large
that two branches appear. This explains the two branches observed in
figures 5 and 6 corresponding to the LOW and (quasi)--vacuum
solutions, respectively (section 4).

Concerning the SMA solution in fig. 4 the main changes arise 
from the bounds on $|g_{ee}|$ and $|g_{e\mu'}|$. 
In the expression of $g_{ee}$ in eq. (10) the contributions of $g_1$ and
$g_2$ may cancel for a phase $\delta \simeq \pi/2$, see fig. 1, as long as
these contributions are of comparable magnitude.
To fulfill this requirement,
a smaller admixture of $g_2$ in $g_{ee}$, as happens for the SMA solution,
has to be compensated by a larger ratio of $g_2/g_1$, corresponding to 
smaller masses $m_1$. This is producing a small hollow in the bound at 
$m_1\sim 5\times 10^{-6}$.

The bound on $|g_{e\mu'}|$ is responsible for the sharp peak at the right 
edge of the excluded region, as can be seen in fig. 2, 
where all Majoron luminosity bounds are shown explicitly.
Here the conservative upper bound on $|g_{e \mu'}|$
is obtained, when $\delta = 0$, corresponding to 
cancellation of the $g_1$ and $g_2$ contributions in the
large $m_1^2$ asymptotics, see eq. (10). Correspondingly the right
border of the excluded region is obtained for $\delta = \pi/2$,
where $|g_{e \mu'}|$ becomes constant for large $m_1$. The 
intersection gives rise to the peak.    
For smaller values of the mixing, as obtained for the SMA solution,
the expression for $|g_{e \mu'}|$ in eq. (10) is fulfilled for corresponding 
larger values of $g_1$. This shifts the exclusion region to the right, 
making it more visible in fig. 4.

\begin{figure}
\hspace*{2cm}
\epsfysize=80mm
\epsfbox{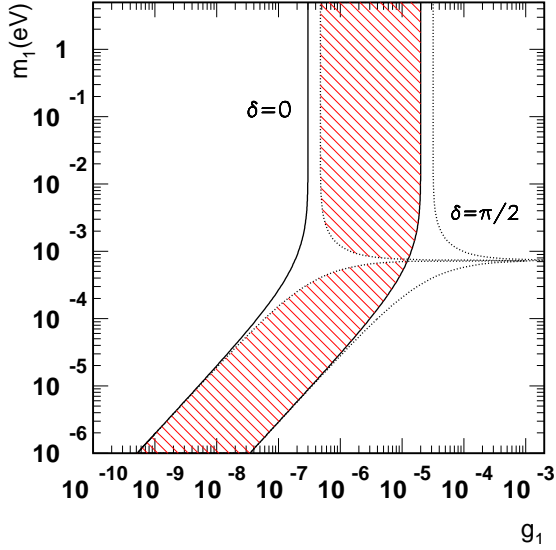}
\caption{Excluded regions from the Majoron luminosity requirement
  $3 \times 10^{-7} < |g_{ee}| < 2 \times 10^{-5}$, for two extreme CP
  cases, $\delta = 0$ (solid lines) and $\delta = \pi/2$ (dotted lines),
  in the $m_1-g_1$ plane.  LMA parameters $\sin^2(2\theta) =
  0.6~,~~\Delta m^2_\odot = 1 \times 10^{-5}$ eV$^2$ are assumed.}
\label{1}
\end{figure}
\begin{figure}
\hspace*{2cm}
\epsfysize=80mm
\epsfbox{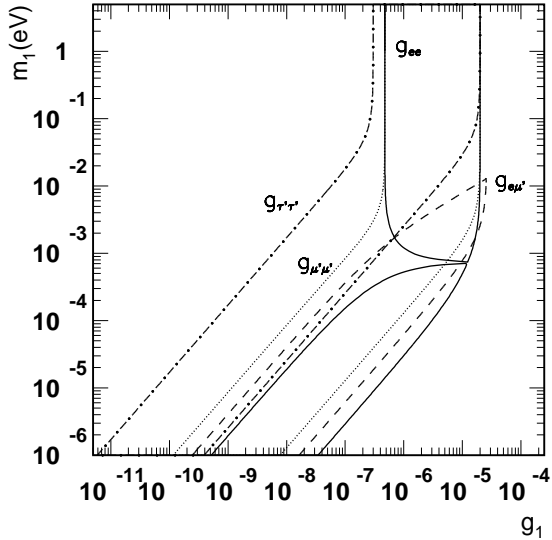}
\caption{Excluded regions independent of CP phase from the Majoron 
  luminosity requirement on $|g_{ee}|$, (solid line), $|g_{e\mu'}|$
  (dashed line), $|g_{\mu'\mu'}|$ (dotted lines) and $|g_{\tau'\tau'}|$
  (dash-dotted line), in the $m_1-g_1$ plane.  LMA parameters
  $\sin^2(2\theta) = 0.6~,~~\Delta m^2_\odot = 1 \times 10^{-5}$
  eV$^2$ are assumed.}
\label{2}
\end{figure}

Before concluding this section we mention the constraints on
Majoron-neutrino coupling parameters which arise from the collapsing
phase.
The idea behind this bound is that a change in the trapped electron
fraction could prevent a successful explosion process.  At the end of
their life massive stars become unstable and, when the iron core
reaches the Chandrasekar limit, they implode.  Once the nuclear
density is reached, a shock wave forms at the core and propagates
outwards, turning the implosion into an explosion. The strength and
propagation of this shock is sensitive to the trapped electron
fraction $Y_{L_e}=Y_e + Y_{\nu_e}$, which can be erased by neutrino
decays $\nu_e \to \overline{\nu}_e + J$. Requiring
$Y_L(t_{bounce})>0.375$ leads to a limit of \cite{kach}
\be
|g_{ee}| < 2  \times  10^{-6}.
\ee
However to the extent that current supernova models do not fully
account for the explosion mechanism this limit should be taken only as
indicative for the moment.

\section{Neutrinoless double beta decay}

The only laboratory experiment, which is competitive with the
supernova bounds, is neutrinoless double beta decay.  This decay
corresponds to two single beta decays occuring in one nucleus and
converts a nucleus (Z,A) into a nucleus (Z+2,A).  Limits on the
Majoron emitting mode
\be
^{A}_{Z}X \to ^A_{Z+2}X + 2 e^- + \phi
\ee
are given by two types of experiments.  In geochemical experiments the
half--life limit is derived from relative abundances of nuclear
isotopes found in the earth \cite{bern}:
\be
|g_{ee}|< 3  \times  10^{-5}.
\ee
However, half--life determinations vary by more than a factor of three.

The best direct laboratory limit (less stringent but more reliable)
from the Heidelberg-Moscow experiment \cite{hdmo} is based on a
likelyhood fit to the continuous electron spectrum:
\be
|g_{ee}|< 8  \times  10^{-5}.
\ee
Future projects such as GENIUS \cite{genius} and EXO \cite{exo} aim at
considerable improvements in the sensitivity.  A very rough estimation
of the sensitivity of GENIUS 1t is based on the background simulation
in \cite{genius}, where a background improvement in the interesting
energy range of a factor $\sim 1000$ has been obtained.  Since in the
Heidelberg-Moscow experiment the Majoron-neutrino coupling bounds are
dominated by the systematical error of the background simulation, a
considerable reduction of the background by a factor of $B$ will
reduce the limit on the Majoron-emitting double beta decay half life
by $\sim \sqrt{B}$ and the coupling constant limit accordingly by
$\sim ^4 \hspace*{-3mm} \sqrt{B}$.  This implies a reach of
sensitivity down to $|g_{ee}| \sim 10^{-5}$, which could bridge most of
the gap existing between the more reliable limits derived from
supernovae.

\section{Discussion and conclusions}

In figures 3 to 6 we present the limits on Majoron-neutrino couplings
in terms of $m_1-g_1$ corresponding to the various solutions of the
solar neutrino problem. In Fig. 7 we display the results for the LMA
solution also in terms of the equivalent $m_2-g_2$ variables. This
representation has been selected for convenience and generality. By
further specifying the underlying model for lepton number violation
one can re-express our results in terms of the lepton number breaking
VEV, which will provide also useful information for model-builders.

Regions which are excluded by supernova arguments are denoted by the
rhombical pattern (obtained from Majoron luminosity) and by the
vertical lines (obtained from the neutrino spectra). Also shown are
the regions excluded from neutrinoless double beta decay (horizontal
lines). The excluded region from Majoron luminosity is a superposition
of the bounds on $\tilde{g}_{11}$, $\tilde{g}_{12}$, $\tilde{g}_{22}$,
and $\tilde{g}_{33}$, where always the most conservative limits for
various CP Majorana phases have been used.  Due to the expressions for
the helicity-flipping neutrino decays the bound obtained from the
neutrino spectra turns out to be independent of the CP phase.  For
neutrinoless double beta decay one can have a cancellation of the
coupling constants $g_1$ and $g_2$. The expected sensitivity of the
GENIUS experiment is shown as a dashed line.  It is easy to see that
GENIUS could be able to bridge almost the whole gap between the
different supernova constraints.  Also an upper bound $m_1<2.3$ eV
from tritium beta decay is displayed.

The limits obtained in this paper apply to the simplest class of
models where neutrino masses arise from the spontaneous violation of
lepton number.  Such Majoron models which cover a wide and attractive
class including both models where the smallness of neutrino
masses follow from a seesaw scheme, as well as those where it arises
from the radiative corrections.


Both neutrinoless double beta decay as well as supernova physics
arguments provide stringent limits on Majoron-neutrino interactions.
In the present work we have discussed these limits and their
translation into the mass basis. Generalizing previous
papers~\cite{kach} we have now taken into account the effect of CP
violating phases, which play a crucial role in the neutrinoless double
beta decay limits.
Depending on the solution of the solar neutrino problem and the
absolute mass scale in the neutrino sector the constraint from the
supernova energy release (Majoron luminosity argument) exludes
Majoron-neutrino couplings in the wide range of $10^{-7}-10^{-5}$.
Upper bounds have been obtained from neutrinoless double beta decay
and the SN87A neutrino spectra. An estimate of the potential of the
future double beta projects such as GENIUS suggests the possibility to
bridge almost the whole gap separating the excluded areas and either
to establish Majorons with couplings around a few $10^{-5}$ or to
restrict neutrino-Majoron couplings down to $10^{-7}$.

Last, but not least, let us mention that the propagation of neutrinos
produced in the solar interior follows essentially the MSW picture,
while any possible effect of decays would happen in vacuo through a
non-diagonal neutrino-majoron coupling which is absent in the simplest
models considered here~\cite{models1}. Even in more complex
models~\cite{mod3,mod4} where such non-diagonal neutrino-majoron
couplings exist in vacuo, one can see that for such small values of
the neutrino-majoron coupling strengths indicated by supernova and
neutrinoless double beta decay, it is rather unlikely that they can
play any role whatsoever in the solar neutrino
problem~\cite{decaysol}.

\vskip 0.2cm
\begin{figure}
\hspace*{2cm}
\epsfysize=80mm
\epsfbox{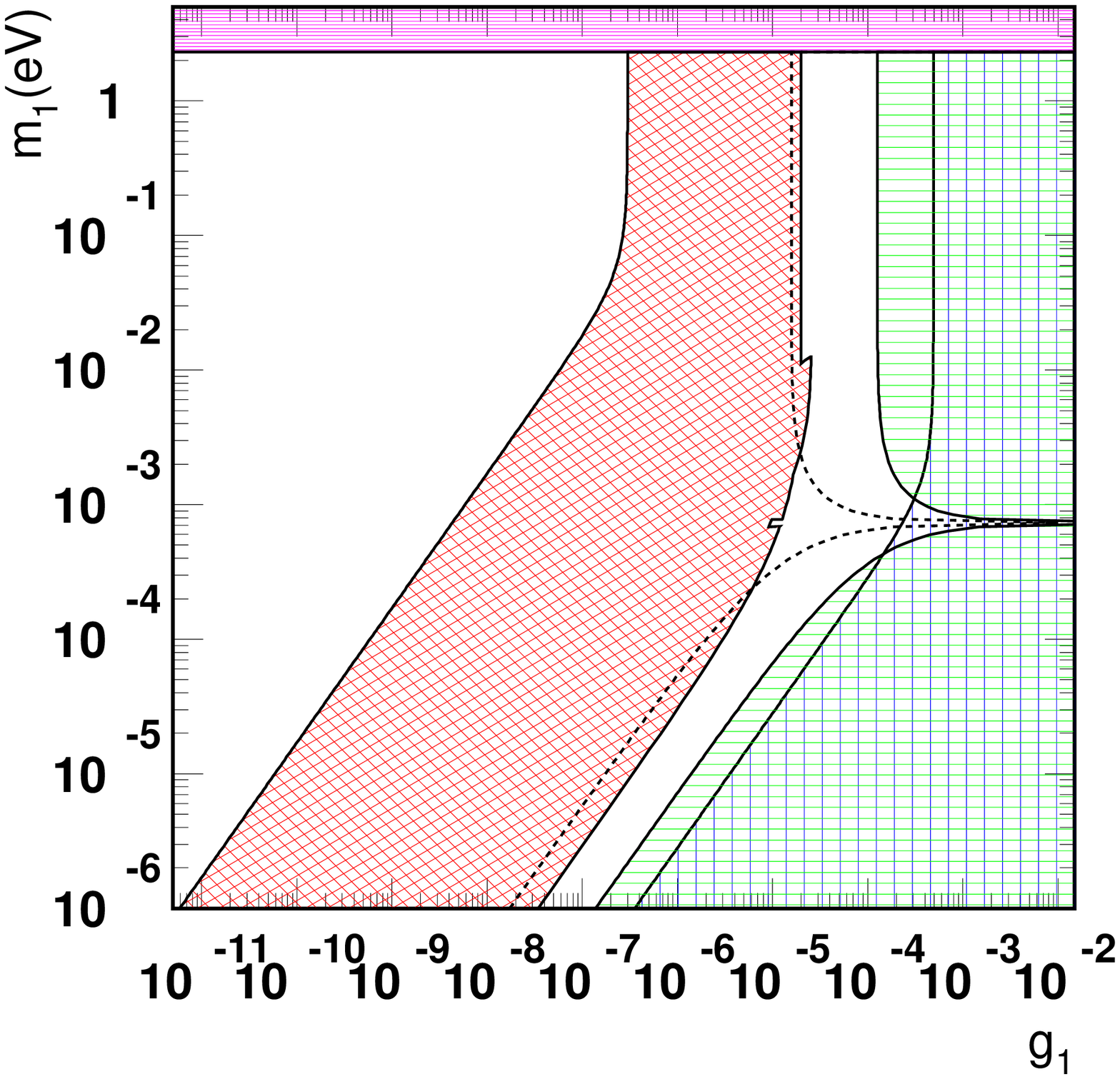}
\caption{Bounds on Majoron models
  in the $g_1-m_1$ plane for the case of the LMA solution ($\Delta
  m^2_{\odot}=10^{-5}$ eV$^2 $ and $ \sin^22\theta_\odot \simeq 0.6$). }
\label{3}
\end{figure}
\begin{figure}
\hspace*{2cm}
\epsfysize=80mm
\epsfbox{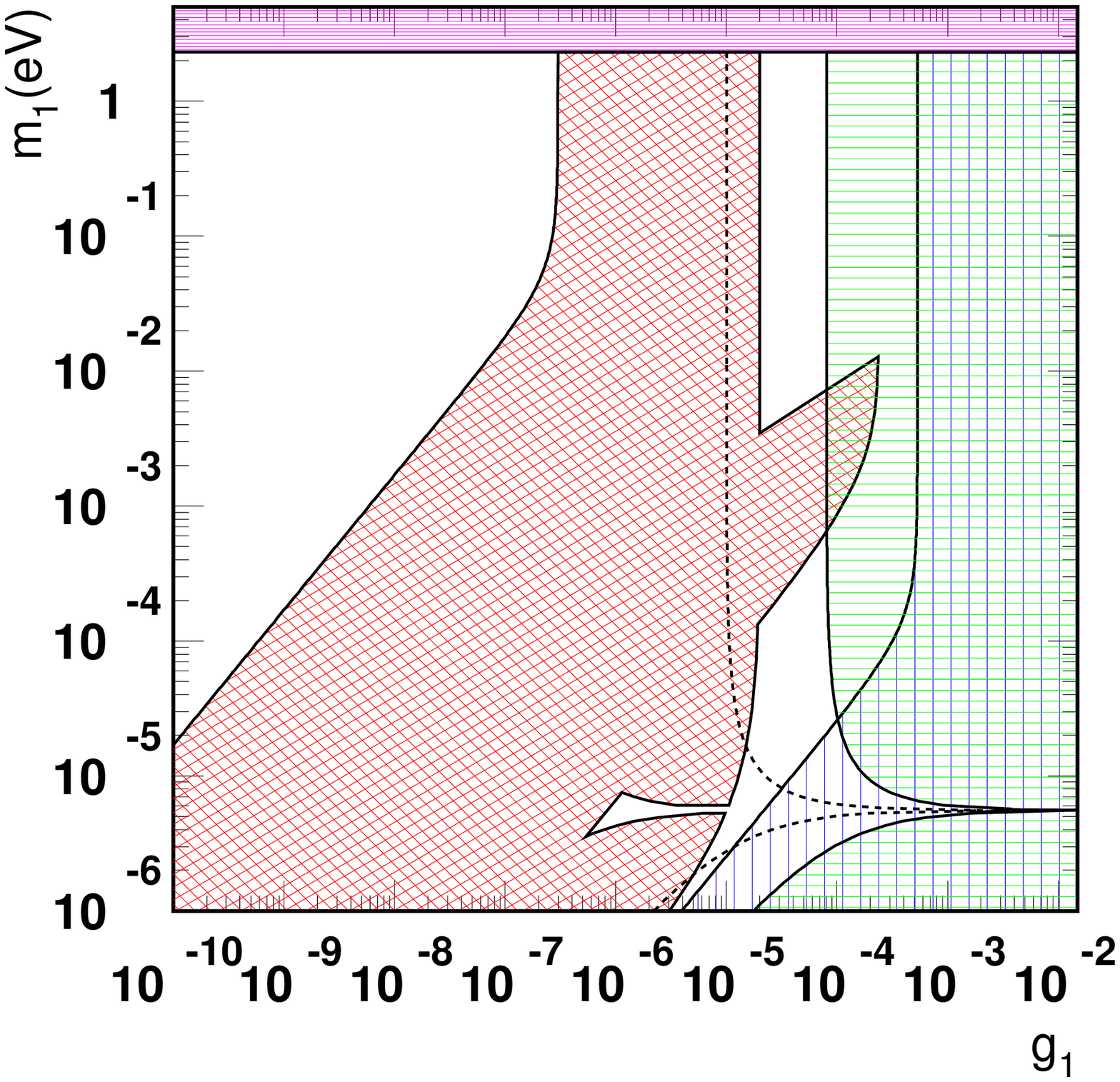}
\caption{Bounds on Majoron models
  in the $g_1-m_1$ plane for the case of the SMA solution ($\Delta
  m^2_{\odot}=10^{-5}$ eV$^2$ and $\sin^22\theta_\odot \simeq 7 \times  10^{-3}$).}
\label{4}
\end{figure}
\begin{figure}
\hspace*{2cm}
\epsfysize=80mm
\epsfbox{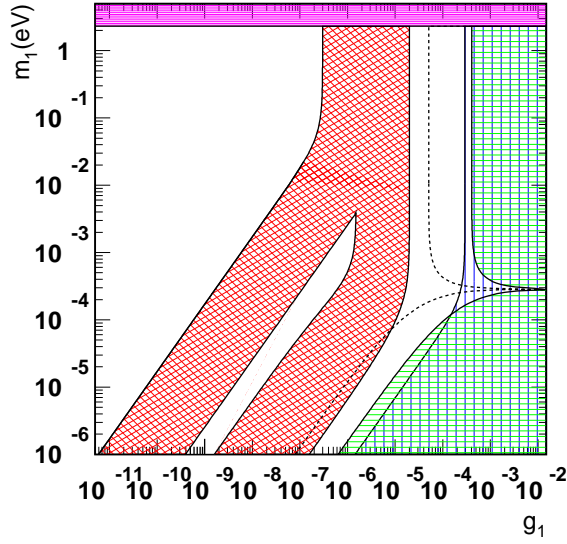}
\caption{Bounds on Majoron models
  in the $g_1-m_1$ plane for the case of the LOW solution ($\Delta
  m^2_{\odot}=10^{-7} eV^2$ and $ \tan^2\theta_\odot \simeq 0.67$).}
\label{5}
\end{figure}
\begin{figure}
\hspace*{2cm}
\epsfysize=80mm
\epsfbox{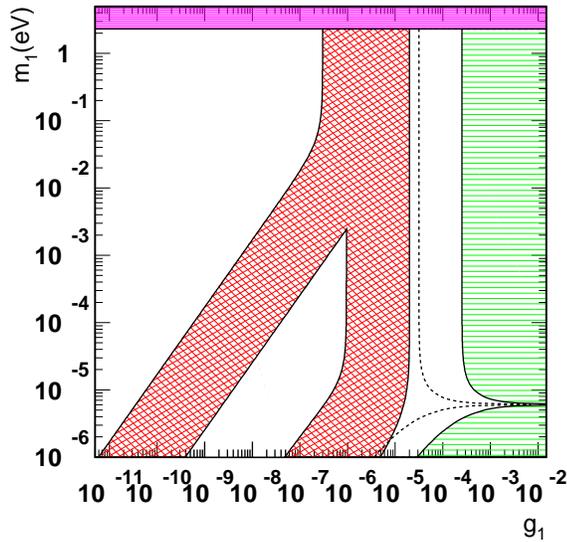}
\caption{Bounds on Majoron models
    in the $g_1-m_1$ plane for the vacuum solution of the solar
    neutrino problem ($\Delta m^2_{\odot}=10^{-10}$ eV$^2$ and
    $\sin^22\theta_\odot\simeq 0.9$. The bounds from neutrino spectra
    don't apply, since already pure neutrino oscillations implied by
    the vacuum solution parameters lead to a contradiction with the
    observed neutrino spectrum of SN1987A. }
\label{6}
\end{figure}
\begin{figure}
\hspace*{2cm}
\epsfysize=80mm
\epsfbox{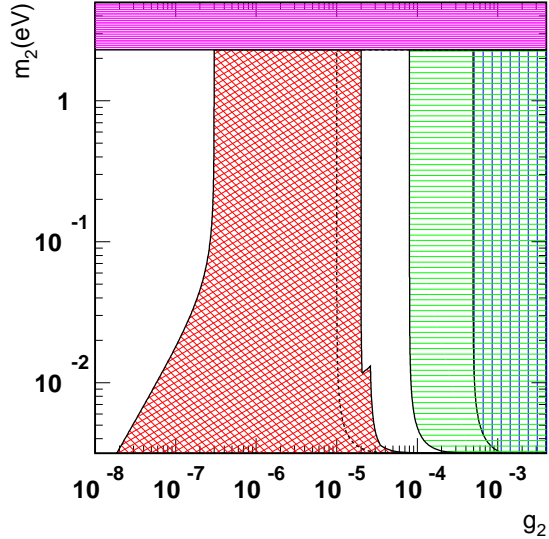}
\caption{
  Bounds on Majoron models expressed in terms ofthe $g_2-m_2$ for the
  case for the LMA solution of the solar neutrino problem.}
\label{7}
\end{figure}


\section*{Acknowledgement}

We thank M. Hirsch for useful discussions.  This work was supported by
Spanish DGICYT under grant PB98-0693, by the European Commission TMR
contract HPRN-CT-2000-00148 and by the European Science Foundation
network grant N.  86. H. P. was supported by TMR contract
ERBFMRX-CT96-0090 and DOE grant no.\ DE-FG05-85ER40226
and R.T. by a grant from the Generalitat Valenciana.

\end{document}